\documentclass{article}
\usepackage{PRIMEarxiv}

\usepackage[hidelinks]{hyperref}
\usepackage{url}
\usepackage{booktabs}
\usepackage{amsfonts}
\usepackage{nicefrac}
\usepackage{microtype}
\usepackage{lipsum}
\usepackage{fancyhdr}
\usepackage{graphicx}
\graphicspath{{./}}
\usepackage{listings}
\usepackage{float}
\usepackage{xcolor}
\usepackage{booktabs}
\usepackage{array}
\usepackage{makecell}
\usepackage{caption}
\usepackage{tikz}
\usetikzlibrary{shapes.geometric, positioning, calc}
\newcolumntype{C}[1]{>{\centering\arraybackslash}m{#1}}

\title{Fingerprint-Driven Automation: Coupling Reconnaissance with POC Verification

}

\author{
  \begin{tabular}[t]{c}
    Hongping Wang \\
    Hainan University \\
    Haikou, China \\
    \texttt{hobinary.wong@gmail.com}
  \end{tabular}
  \hfill
  \begin{tabular}[t]{c}
    Xiaoqi Li \\
    Hainan University \\
    Haikou, China \\
    \texttt{csxqli@ieee.org}
  \end{tabular}
}

\begin{document}
\maketitle

\begin{abstract}
In the field of network security confrontation, reconnaissance is the first and most critical step. Accurate, efficient, and comprehensive reconnaissance can help network security workers more fully understand the target's current state, identify potential weaknesses, and formulate a targeted attack strategy. However, there are some problems in the existing tools on the market, such as low accuracy of the collected information, lack of concealment, time-consuming, low or high integration of tools, which makes them difficult to start using. With the continuous development of technology, systems need to continuously upgrade their technologies and strategies, and expand new functions. The scalability of tool functions is also one of the factors that need to be taken into consideration. These problems in the practice of network security have a serious impact on user experience and work efficiency. In view of this, a more user-friendly, more automated reconnaissance and vulnerability verification tool is designed. This paper proposes to develop a highly automated reconnaissance and vulnerability verification tool. This paper presents the tool functions and conducts the basic design. Finally, we implement and test all the functions specifically to verify the feasibility of the tool functions and highlight the advantages of this tool in terms of automation, retrieval, and use in network security operations. Through the test of the online security simulation range, the results show that this tool can complete the automatic reconnaissance and in-depth data processing of specific targets, and conduct automatic vulnerability detection and verification.

\end{abstract}
% keywords can be removed

\section{Introduction}

In the field of cybersecurity, there are numerous tools available for security testing\cite{alhamed2023systematic}, which can be broadly categorized into three types: large scale automated vulnerability scanning systems (such as AWVS, which specializes in web application vulnerability scanning)\cite{harzevili2025systematic}\cite{harzevili2023survey}\cite{zhang2022efficiency}, tools designed for specific functions (such as Dirsearch, which specially performs directory scanning), and vulnerability detection systems tailored to specific frameworks (such as WPscan, which is specifically designed to identify security risks in WordPress websites\cite{wu2025exploring}\cite{zhou2025blockchain}\cite{zhu2024sybil}). While the first category of large-scale systems offers a vast number of proof-of-concept (PoC) tests and comprehensive vulnerability detection, they require significant scanning time, often charge for certain functions, and present a steep learning curve for beginners\cite{amankwah2020empirical}\cite{koman2025scanme}. The latter two categories, having been developed for specific functions, have a narrower scope of application\cite{mazurczyk2021cyber}.

Some tools suffer from design flaws or errors, for example, the directory scanning tool Dirsearch typically relies solely on HTTP response codes to determine page accessibility. If a site uses a custom error page or has incorrect server configuration, and the error page returns a 200 response code, the tool may misjudge the situation and return an invalid page. Additionally, many security tools are only semi-automated. When gathering basic information, users often have to manually inspect front-end code, JavaScript files, or HTML comments to extract sensitive data\cite{ferrara2012web}\cite{sarumathi2024benchmark}\cite{essien2023ethics}. This process is time-consuming and error-prone, significantly reducing the efficiency of security work. Furthermore, the logic behind some service identification features is overly simplistic, relying solely on signature ports or response codes for inference. This lack of depth and comprehensiveness makes it difficult to adapt to the complex characteristics of modern web applications. At the same time, many security systems have shortcomings in terms of functionality maintenance and expansion\cite{rahman2019systematic}\cite{leite2020survey}. They were not designed to allow users to customize and extend tool functions, and when developers later abandon maintenance, the features fail to meet the demands of technological advancements, ultimately rendering the tool unusable.

Therefore, to meet the practical needs of cybersecurity professionals, it is necessary to design and propose a security tool with good maintainability and scalability\cite{ivancic2019robotic}\cite{wiklund2017impediments}, ease of use, lightweight design, accurate results, and a high degree of automation\cite{bianou2024pentest}\cite{xu2025large}. This paper analyzes the shortcomings of traditional cybersecurity tools in practical applications. Based on the actual needs of security professionals and leveraging Python's extensive third-party libraries\cite{fang2024llm}\cite{sheng2026llms}\cite{bu2025smartbugbert}\cite{fang2024novelty}\cite{paramitha2023technical}, we design and develop a reconnaissance\cite{yadav2023open}\cite{evangelista2021systematic} and vulnerability verification tool for the cybersecurity reconnaissance phase that delivers accurate results, is highly automated, highly scalable, and lightweight yet comprehensive.

This paper makes the following contributions:
\begin{enumerate}
    \item[(1)] We present a novel automated penetration testing framework that seamlessly auto-links the two separate and traditionally disjoint phases of reconnaissance and vulnerability verification into a security testing closed-loop pipeline. 
    \item[(2)] We propose a fingerprint-based nexus mechanism as the main technical linkage of the two phases. During reconnaissance, we capture precise web fingerprints and dynamically map the fingerprints to flexibly invoke highly faithful PoC scripts, so as to make penetration testing targeted, avoid scanning blindness, and reduce network traffic overhead. 
    \item[(3)] Instead of just using synthetic experiments, we do intensive evaluations in real scenarios on some authentic Capture the Flag (CTF) platform environments. These platform environments have complex and real vulnerable chains and authentic service responses. They allow the tool to be practically feasible, robust, and practical in real-world offensive security scenarios.
\end{enumerate}
\section{Related Work}
The existing network security testing tools mainly cover three categories: large-scale automated vulnerability scanning systems, specific function-oriented special tools, and vulnerability detection systems developed for specific frameworks\cite{li2024interaction}\cite{shen2023intellicon}\cite{li2025penetrating}. The first type of tool, represented by AWVS, although it provides a PoC test case library with a wide range of coverage and can detect various types of security vulnerabilities, its scanning process takes a long time, advanced functions usually need to be paid for use, and the interface and operation logic have high learning costs for beginners \cite{amankwah2020empirical}\cite{koman2025scanme}. Comparatively speaking, tools such as Dirsearch, a directory scanning tool, or WPscan, a special scanner for WordPress, have advantages in terms of resource occupation and ease of operation, but at the cost of significantly narrowing the scope of application, they often only cover a specific link in the security testing workflow and are difficult to meet the comprehensive detection requirements in complex scenarios.

In addition to the above classification characteristics, there are some common problems with various tools. In terms of detection accuracy, the design mechanism of some tools itself introduces bias. Dirsearch judges page accessibility only based on the HTTP response status code, which often leads to false judgments in actual tests. For example, when the target website is configured with a custom error page, and the page returns a 200 status code, Dirsearch will mark the path that does not actually exist as a valid resource. Testers will conduct subsequent analysis based on these results, which is inevitably misleading. In terms of the degree of automation, a considerable number of security tools only achieve semi-automatic workflow. The reconnaissance phase still relies on manual intervention. Users need to check the front-end code, JavaScript files, and HTML comments one by one to extract sensitive information that may be included \cite{ferrara2012web}\cite{sarumathi2024benchmark}, such as debugging interfaces, temporary credentials, or internal paths left by developers. This process is time-consuming and error-prone, especially in pages with large code volume, the risk of omission in manual review increases significantly. There is also a tendency to simplify the logic at the service identification level. Currently, many tools only rely on the default port number or the Server field in the response header to determine the type of back-end components. However, the technology stack of modern Web applications is becoming increasingly complex, and container deployment, content distribution network(CDN) acceleration, and the widespread application of reverse proxy make this single-dimension identification strategy difficult to work with and often misjudge or omit the actual running service types. In addition, the lack of tool scalability cannot be ignored. Most tools do not provide plug-in mechanisms or API interfaces, and users cannot add new detection modules or adjust existing logic according to their own needs. Once the original development team stops maintenance, these tools will quickly become disconnected from the new vulnerability types, and their practicability will decline \cite{rahman2019systematic}\cite{leite2020survey}\cite{kahlon2026systematic}.

The more prominent problem is the fracture at the architecture level. The two stages of reconnaissance and vulnerability verification are usually designed as independent modules in the existing toolset, with independent input/output formats and execution logic. This means that security analysts must manually sort out all kinds of clues output in the early reconnaissance stage, screen out suspicious points, and then select appropriate verification scripts to perform the next step. This manual bridging process not only slows down the overall test rhythm, but also introduces the possibility of additional human errors, it is not uncommon for key clues to be missed or misread in the transmission process. In recent years, although some studies have tried to introduce large language models into the automated penetration test process \cite{fang2024llm}\cite{sheng2026llms}\cite{li2025no}, or use the agent framework to open the path between reconnaissance and verification, these schemes generally face the problem of high computing overhead and lack a lightweight middle layer mechanism to efficiently map the reconnaissance results to executable verification actions.

In view of the above shortcomings, this paper proposes an automation framework based on fingerprint drive. The core design idea of this framework is to closely couple reconnaissance and verification to form a closed-loop security test process. Specifically, the contribution of this paper is reflected in three levels. First, we have built an end-to-end automated pipeline. The results from the reconnaissance phase are no longer handed over to manual processing as isolated data sets, but directly flow into the verification phase to trigger the corresponding detection tasks, thus eliminating the phase connection in the traditional workflow that relies on people. Secondly, we designed a fingerprint-based association mechanism. During the reconnaissance process, the tool extracts website fingerprints from multiple dimensions such as response content, response headers, and static resource characteristics, covering technology stack type, server version, and application framework information, and then dynamically matches and calls the corresponding PoC verification script based on this fingerprint information. This strategy makes the testing process more focused, avoids the network resource consumption caused by blindly traversing all known vulnerabilities, and also improves detection efficiency. Thirdly, we verified the tool in the real CTF competition environment. The test cases include a variety of target applications with real vulnerability chains and various abnormal service response scenarios. The experimental results show that the tool can stably complete the closed-loop process of detection verification when facing different target instances, and the rates of false positives and false negatives are controlled within an acceptable range, which indicates that the tool has good feasibility and robustness in actual combat attack and defense scenarios.

\section{Design}
\subsection{Architecture Design}
The overall functional architecture of this system aims to build an end-to-end automated security assessment pipeline. The architecture mainly consists of three parts: a multi-dimensional basic information reconnaissance module, a fingerprint-driven scanning and identification module, and a heterogeneous data aggregation and processing module\cite{difrancesco2019architecting}. These three functional modules maintain a high degree of independence in the implementation of the underlying code to facilitate subsequent independent iteration and plug-in expansion, however, at the logic and data flow level, synchronous collaboration is achieved through close series connection of standardized internal data interfaces. As the starting point of the assembly line, the reconnaissance module is responsible for non-intrusive asset mapping and basic information extraction of the target to build an initial target portrait, the scanning and identification module serves as the core hub connecting the previous and the following, relying on the portraits provided by pre-reconnaissance to accurately perform fingerprint matching and directional detection, effectively avoiding the high time cost and risk of false positives brought by traditional blind scanning, The data processing module is responsible for deep cleaning and structural processing of multi-source and heterogeneous detection echo data, providing effective data support for subsequent PoC triggering and vulnerability verification. The seamless connection of these three modules breaks the data island between traditional security tools and ensures the efficiency and robustness of the entire automated penetration process. The specific functional topology is shown in Figure 1.
\begin{figure}[H]
    \centering
    \includegraphics[height=0.65\linewidth]{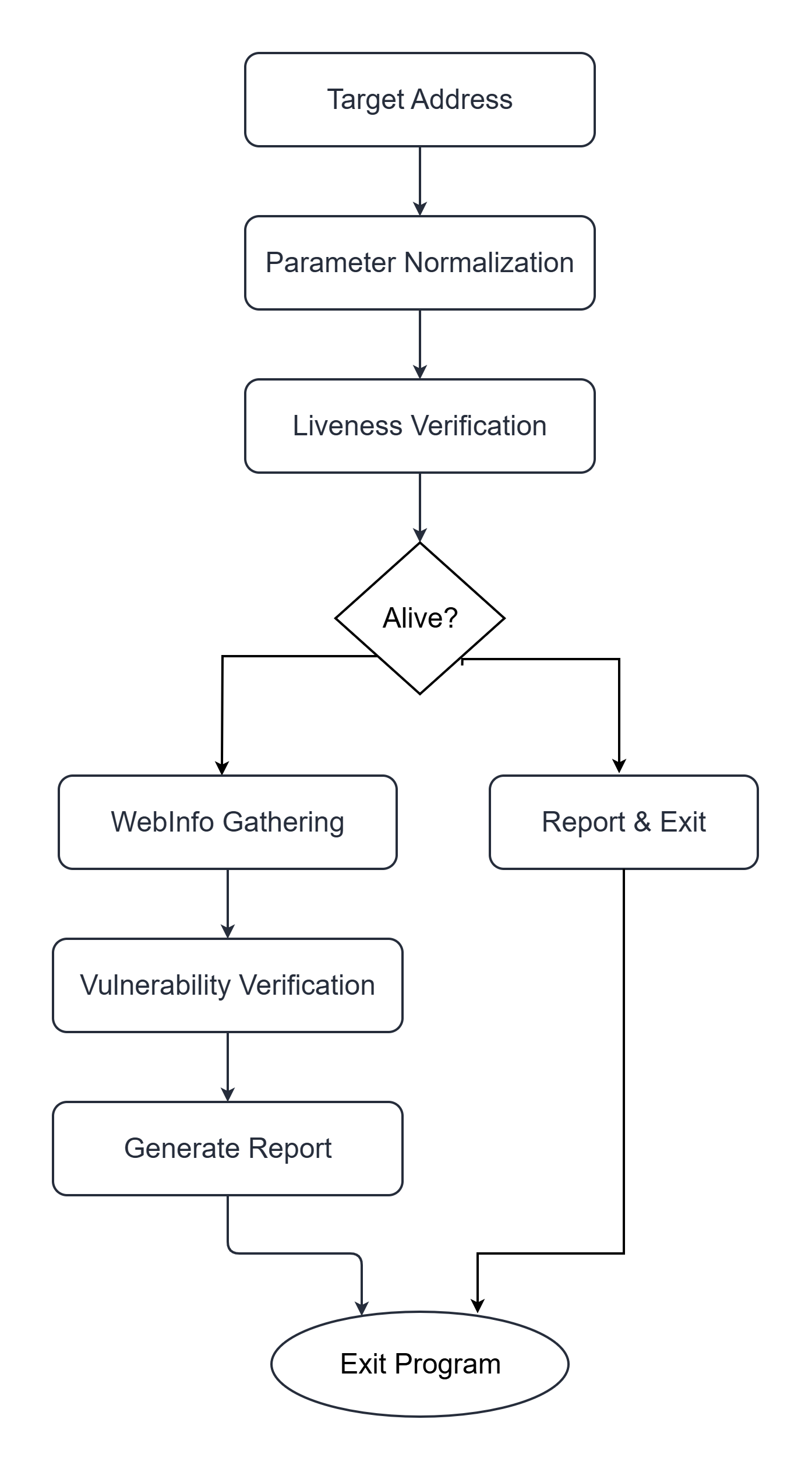}
    \renewcommand{\thefigure}{1}
    \caption{Functional Diagram}
\end{figure}
\subsection{Functional Design}
The execution flow of the entire tool consists of the following steps: collection of basic information, vulnerability information matching, and generation of the scan report. The vulnerability information matching process requires the use of the necessary basic information parameters collected earlier, while the report includes two sections: basic information data and details regarding the presence of vulnerabilities. Figure 2 illustrates the execution flow of each function during system operation.
\begin{figure}[H]
    \centering
    \includegraphics[width=1.0\linewidth]{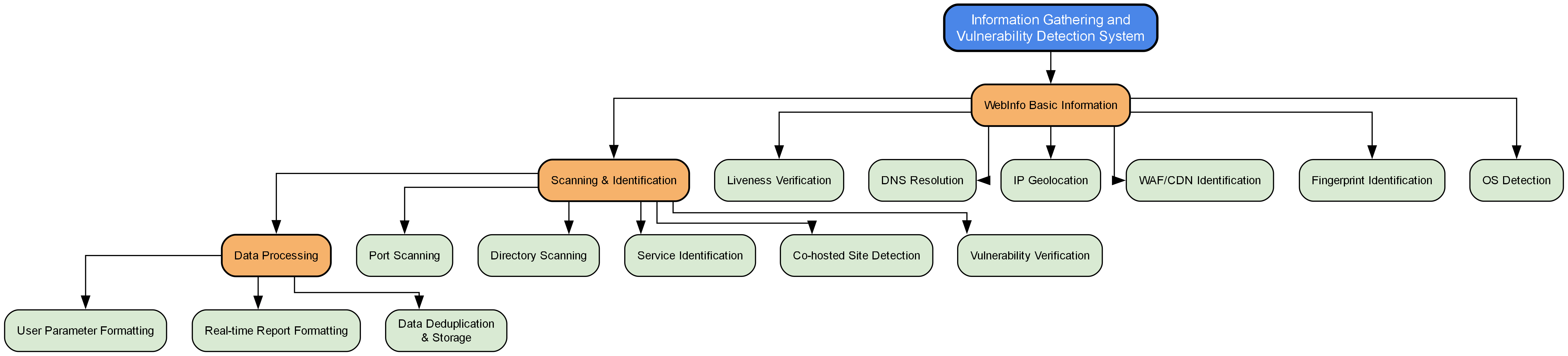}
    \renewcommand{\thefigure}{2}
    \caption{Function Flowchart }
\end{figure}
\subsubsection{User Interface}
The user interface is the initial presentation of the system to users. Whether it is a terminal command-line interface or a graphical web interface, it should be simple and clear, highlighting the tool's core functions. The tool offers two interface options, a terminal command-line interface and a graphical web interface, to meet the needs of different users.

Deepscan's terminal command-line interface (CLI) strictly follows the design principles of minimalism and operational efficiency, and is designed to provide a smooth and intuitive interactive experience for security assessors. In the system initialization phase, the tool first displays the system name by rendering the customized ASCII art font. This design not only gives the tool professional visual beauty and brand recognition, but also provides users with a clear visual anchor in the complex terminal data stream, thus helping the operator quickly establish the working context. Followed by a concise summary of the core parameters and basic usage of the system, this pre-interactive guidance effectively reduces the learning threshold of the tool and ensures that users can quickly master key instructions. More importantly, before any automatic scanning task is formally triggered, the interface will explicitly enumerate and echo the currently active global configuration parameters. This pre execution state confirmation mechanism not only ensures the transparency of the scanning process to the greatest extent, enabling users to visually verify the execution boundary and detection logic of the task, but also effectively avoids invalid network overhead or potential compliance risks caused by parameter mismatches, which further strengthens the precision and controllability of the security test process while improving the overall operational efficiency.

To enhance user experience and ease of use, Deepscan has added a web-based graphical interface, allowing users to access and operate the tool directly via a browser. The web interface utilizes a modern design style and  the Bootstrap framework to achieve a responsive layout, ensuring optimal display across various devices.

\subsubsection{Webinfo Information}
The functional design of the webinfo module aims to carry out a comprehensive network environment detection of the target web assets and provide more verification perspectives for subsequent accurate vulnerability verification. The module first obtains the real IP address of the target through deep domain name resolution, and constructs the network space geographic mapping combined with IP geographic location technology to establish the real network boundary of the target. The system further performs multi-dimensional feature extraction. It not only accurately captures the surface metadata such as the web page title, but also deeply analyzes the HTTP response message to obtain high-value web fingerprint data, middleware server types, and the details of the underlying operating system, so as to fully obtain the back-end technology ecology of the target. For the complex defense system in the modern web architecture, the webinfo module integrates the active network boundary detection mechanism, which can accurately determine whether the target front-end has deployed CDN or web application firewall (WAF), and further identify the specific WAF manufacturer characteristics and protection configuration. This series of in-depth detection logic from the outside to the inside, from the basic assets to the defense mechanism, not only effectively avoided the false positives and failures caused by CDN cache interference or WAF dynamic interception in the subsequent vulnerability verification stage, but also perfectly matched the core idea of "fingerprint driven" in this system, laying a solid reconnaissance foundation for the smooth progress of automated penetration testing.
\subsubsection{Port/Directory Scanning}
Determine which ports are open and identify specific services based on characteristic ports, while utilizing multithreading for acceleration. Additionally, port spoofing must be considered, if more than 30 ports are open, port spoofing is highly probable.

In the directory scanning module of automated penetration testing, to achieve efficient and accurate asset discovery, we need to comprehensively consider the dynamic generation of scanning dictionaries, the scheduling mechanism of high concurrent requests, and complex false positive suppression strategies. The core is to conduct a multi-dimensional in-depth analysis of the HTTP response content to confirm the true existence of the target file or directory. In order to effectively peel off the interference information returned by the server and greatly reduce the false positive rate, at the basic protocol response level, the system will perform strict status code verification to directly eliminate common server error responses, and synchronously review the HTTP response header and content type, so as to accurately intercept invalid probes that carry custom error IDs or MIME types that are seriously inconsistent with expectations. At the semantic analysis level of the response body, the module introduces a feature matching algorithm for the "soft 404" phenomenon. By extracting and comparing the specific DOM structure or feature text in the general or custom 404 page, it thoroughly filters out those pseudo-live directories that return a 200 status code but have invalid actual content. In view of the complex routing behavior prevalent in modern web applications, the system will deeply track and verify the legitimacy and accessibility of the redirect target URL to prevent invalid jump interference caused by malicious configuration or interception by protective devices (such as WAF). In addition, the system also has a built-in dynamic special rule engine based on file attributes, which can perform customized fingerprint verification logic for different types of sensitive files (such as backup files, configuration files, or specific script suffixes). By implementing the above multi-layer comprehensive research and judgment strategy, not only the throughput efficiency of concurrent scanning is maximized, but also the extremely high accuracy and practical reliability of directory enumeration results are fundamentally guaranteed.
\subsubsection{Service Fingerprinting}
Using the open-source Webeyes and Wappalyzer fingerprint libraries, crawl the target's pages, response headers, the `src` attribute of `script` tags, and meta information, then perform regular expression matching against the fingerprint libraries to complete fingerprint identification.
\subsubsection{Proof Of Concept}
The tool employs fingerprinting technology to identify web fingerprints. During the reconnaissance phase, it collects specific field information from web applications, such as page keywords, special files and paths, based on URLs entered by the user. In the web fingerprinting phase, a fingerprint database is established to collect feature information and build a fingerprint feature repository. Simultaneously, the tool gathers feature information from the target web application and compares it with the data in the fingerprint feature repository to perform web fingerprinting. The fingerprint information in the fingerprint database is collected and synthesized from platforms such as Wappalyzer and FOFA, and is stored in JSON format. Fingerprint recognition technology is a critical step in the penetration testing process, helping us quickly formulate penetration strategies\cite{aydos2022security}. Based on the WebInfo and fingerprint information collected earlier, we determine whether the conditions for executing the PoC script are met. A dedicated verification function must be written to accomplish this task, thereby achieving the effectiveness of the PoC.
\subsubsection{Database Design}
In order to realize the efficient access management of data in the automatic penetration process, the system designs a structured core data model, which mainly covers four progressive key entities. First of all, webinfo (WEB basic information) entity as the cornerstone of building the target image is responsible for persisting the underlying technology stack and core fingerprint features of the target site, Secondly, ports entity accurately depicts the network exposure of the target, and records the open port status and its associated background service types in detail, On this basis, urls (resource path) entity is further extended to the application layer, which is specially used to map and store sensitive paths and key files found by deep directory enumeration, Finally, all security defects successfully verified by PoC scripts will be aggregated and written into the vulns (vulnerability information) entity in a standardized manner. These four entities are logically closely linked, completely mapping the data link from asset mapping, service identification to vulnerability verification, providing a solid and reliable database for the fingerprint-driven mechanism and automated decision-making of the system.

The relationships between tables are presented as an entity-attribute diagram. A WebInfo entry represents a target website, and WebInfo maintains a 1:N relationship with all other entities, meaning each website may have N entries for port information, directory information, crawler results, and vulnerability information. 

\begin{figure}[H]
    \centering
    \label{fig:er_diagram}
	\resizebox{0.95\textwidth}{!}{
		\begin{tikzpicture}[
			node distance=1.2cm,
			entity/.style={draw, rectangle, minimum height=0.6cm, minimum width=1.2cm, font=\bfseries, align=center},
			attribute/.style={draw, ellipse, minimum height=0.5cm, minimum width=1cm, align=center, font=\small},
			relation/.style={draw, diamond, aspect=1.5, inner sep=1pt},
			edge_label/.style={font=\small, inner sep=2pt}
		]
		
		\node[entity] (webinfo) {WebInfo};
		
		\node[attribute, above=0.6cm of webinfo] (os) {OS};
		\node[attribute, above right=0.6cm and 1.2cm of webinfo] (server) {Server};
		\node[attribute, right=1.2cm of webinfo] (ip) {IP Location};
		\node[attribute, below right=0.6cm and 1.2cm of webinfo] (md5) {MD5};
		\node[attribute, below=0.6cm of webinfo] (app) {Application};
		\node[attribute, below left=0.6cm and 1.2cm of webinfo] (time_w) {Time};
		\node[attribute, left=1.2cm of webinfo] (fw) {Firewall};
		\node[attribute, above left=0.6cm and 1.2cm of webinfo] (dom) {Domain};
		
		\foreach \n in {os, server, ip, md5, app, time_w, fw, dom} \draw (webinfo) -- (\n);

		\node[relation, above left=1.5cm and 2.5cm of webinfo] (rel_dir) {Scan};
		\node[edge_label, above=0.1cm of rel_dir] {N};
		\node[edge_label, right=0.1cm of rel_dir] {1};
		\node[entity, above left=1.2cm and 1.5cm of rel_dir] (directory) {Directory};
		\node[attribute, above=0.6cm of directory] (d_type) {Type};
		\node[attribute, right=0.6cm of directory] (d_dir) {Directory};
		\node[attribute, below right=0.6cm and 0.1cm of directory] (d_time) {Time};
		\node[attribute, below=0.6cm of directory] (d_md5) {MD5};
		\node[attribute, below left=0.6cm and 0.1cm of directory] (d_len) {Resp. Length};
		\node[attribute, left=0.6cm of directory] (d_code) {Resp. Code};
		\node[attribute, above left=0.6cm and 0.1cm of directory] (d_page) {Page Name};
		\node[attribute, above=0.6cm and 1.2cm of d_page] (d_dom) {Domain};
		
		\draw (webinfo) -- (rel_dir) -- (directory);
		\foreach \n in {d_type, d_dir, d_time, d_md5, d_len, d_code, d_page, d_dom} \draw (directory) -- (\n);

		\node[relation, above right=1.5cm and 2.5cm of webinfo] (rel_port) {Scan};
		\node[edge_label, above=0.1cm of rel_port] {N};
		\node[edge_label, left=0.1cm of rel_port] {1};
		\node[entity, above right=1.2cm and 1.5cm of rel_port] (port) {Port};
		\node[attribute, above left=0.6cm and 0.1cm of port] (p_serv) {Service};
		\node[attribute, above right=0.6cm and 0.1cm of port] (p_bann) {Banner};
		\node[attribute, right=0.6cm of port] (p_time) {Time};
		\node[attribute, below right=0.6cm and 0.1cm of port] (p_ip) {IP Address};
		\node[attribute, below=0.6cm of port] (p_prt) {Port};
		\node[attribute, below left=0.6cm and 0.1cm of port] (p_md5) {MD5};
		
		\draw (webinfo) -- (rel_port) -- (port);
		\foreach \n in {p_serv, p_bann, p_time, p_ip, p_prt, p_md5} \draw (port) -- (\n);

		\node[relation, below left=1.5cm and 2.5cm of webinfo] (rel_vuln) {Detect};
		\node[edge_label, below=0.1cm of rel_vuln] {N};
		\node[edge_label, right=0.1cm of rel_vuln] {1};
		\node[entity, below left=1.2cm and 1.5cm of rel_vuln] (vuln) {Vulnerability};
		\node[attribute, above left=0.6cm and 0.1cm of vuln] (v_time) {Time};
		\node[attribute, left=0.6cm of vuln] (v_dom) {Domain};
		\node[attribute, below left=0.6cm and 0.1cm of vuln] (v_md5) {MD5};
		\node[attribute, below right=0.6cm and 0.1cm of vuln] (v_info) {Vuln Info};
		
		\draw (webinfo) -- (rel_vuln) -- (vuln);
		\foreach \n in {v_time, v_dom, v_md5, v_info} \draw (vuln) -- (\n);

		\node[relation, below right=1.5cm and 2.5cm of webinfo] (rel_crawl) {Mine};
		\node[edge_label, below=0.1cm of rel_crawl] {N};
		\node[edge_label, left=0.1cm of rel_crawl] {1};
		\node[entity, below right=1.2cm and 1.5cm of rel_crawl] (crawler) {Crawler};
		\node[attribute, above right=0.6cm and 0.1cm of crawler] (c_time) {Time};
		\node[attribute, right=0.6cm of crawler] (c_dom) {Domain};
		\node[attribute, below right=0.6cm and 0.1cm of crawler] (c_type) {Info Type};
		\node[attribute, below=0.6cm of crawler] (c_md5) {MD5};
		\node[attribute, below left=0.6cm and 0.1cm of crawler] (c_leak) {Leaked Info};
		
		\draw (webinfo) -- (rel_crawl) -- (crawler);
		\foreach \n in {c_time, c_dom, c_type, c_md5, c_leak} \draw (crawler) -- (\n);
		
		\end{tikzpicture}
	}
    \renewcommand{\thefigure}{3}
    \caption{Entity-Relationship Diagram of the System}
\end{figure}
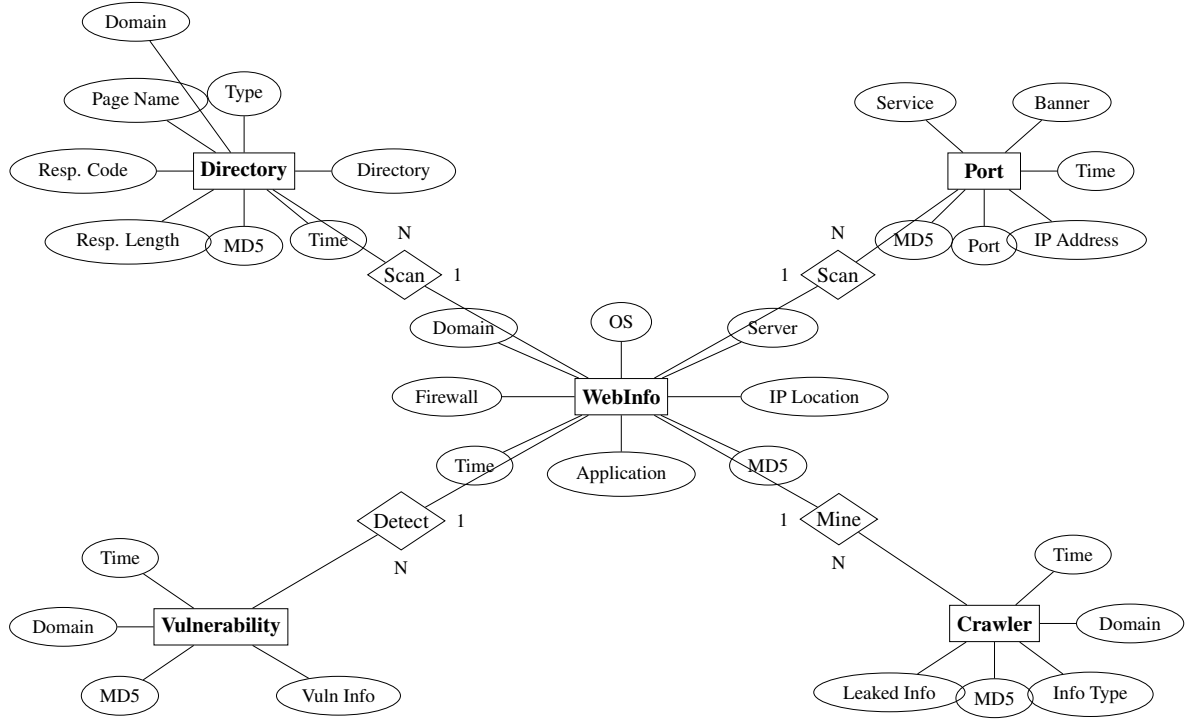

(1) Website Information Table (WebInfo): Primarily records the technical stack information of the target site, including domain name, server type, WAF, open services, and other details. The design of the Website Information Table is shown in Table 1.
\begin{table}[H]
  \centering
  \renewcommand{\thetable}{1}
  \caption{WebInfo Table}
  \label{tab:webinfo}
  
  \begin{tabular}{C{2cm} C{4cm} C{9cm}} 
    \toprule
    \textbf{Field Name} & \textbf{Data Type} & \textbf{Description} \\
    \midrule
    id & INTEGER & Auto-increment primary key \\
    time & TIMESTAMP & Timestamp \\
    domain & VARCHAR(255) & Domain name \\
    waf & VARCHAR(100) & WAF type \\
    apps & TEXT & Application software used \\
    server & VARCHAR(100) & Server type \\
    ipaddr & VARCHAR(45) & IP address \\
    os & VARCHAR(100) & Operating system \\
    md5 & CHAR(32) & Unique identifier (MD5 hash) \\
    \bottomrule
  \end{tabular}
\end{table}

\vspace{1em}
(2) Port Information Table (Ports): Primarily records the open port information of the target site, including IP address, open ports, and services corresponding to specific ports. The design of the Port Information Table is shown in Table 2.
\begin{table}[H]
  \centering
  \renewcommand{\thetable}{2}
  \caption{Ports Table}
  \label{tab:ports}
  
  \begin{tabular}{C{2cm} C{4cm} C{9cm}} 
    \toprule
    \textbf{Field Name} & \textbf{Data Type} & \textbf{Description} \\
    \midrule
    id & INTEGER & Auto-increment primary key \\
    time & TIMESTAMP & Timestamp \\
    ipaddr & VARCHAR(45) & IP address \\
    service & VARCHAR(100) & Service type \\
    port & INT & Port number \\
    banner & TEXT & Service Banner \\
    md5 & CHAR(32) & Unique identifier (MD5 hash) \\
    \bottomrule
  \end{tabular}
\end{table}
(3) Directory Information Table (URLs): Primarily records valid directory and file information identified after scanning the target site, including domain name, page title, full web address, and response content. The design of the Directory Information Table is shown in Table 3.
\begin{table}[H]
  \centering
  \renewcommand{\thetable}{3}
  \caption{Urls Table}
  \label{tab:urls}
  
  \begin{tabular}{C{2cm} C{4cm} C{9cm}} 
    \toprule
    \textbf{Field Name} & \textbf{Data Type} & \textbf{Description} \\
    \midrule
    id & INTEGER & Auto-increment primary key \\
    time & TIMESTAMP & Timestamp \\
    domain & VARCHAR(255) & Domain name \\
    title & VARCHAR(255) & Page Title \\
    url & VARCHAR(2048) & Complete URL \\
    contype & VARCHAR(100) & Content Type \\
    rsp\_len & INT & Response Length \\
    rsp\_code & SMALLINT & Response Code \\
    md5 & CHAR(32) & Unique identifier (MD5 hash) \\
    \bottomrule
  \end{tabular}
\end{table}
(4) Vulnerability Information Table (Vulns): Primarily records verified valid vulnerability information for the target site, including domain name and vulnerability details. The design of the Vulnerability Information Table is shown in Table 4.
\begin{table}[H]
  \centering
  \renewcommand{\thetable}{4}
  \caption{Vulns Table}
  \label{tab:vulns}
  
  \begin{tabular}{C{2cm} C{4cm} C{9cm}} 
    \toprule
    \textbf{Field Name} & \textbf{Data Type} & \textbf{Description} \\
    \midrule
    id & INTEGER & Auto-increment primary key \\
    time & TIMESTAMP & Timestamp \\
    domain & VARCHAR(255) & Domain name \\
    vuln & VARCHAR(512) & Vulnerability Description \\
    md5 & CHAR(32) & Unique identifier (MD5 hash) \\
    \bottomrule
  \end{tabular}
\end{table}
\subsubsection{Report Generation}
The report generation module of DeepScan aims to achieve efficient processing and multi-dimensional visualization of scanned data through the underlying database interaction mechanism. At the data persistence level, the system uses the lightweight SQLite3 as the storage engine and encapsulates the custom database operation class 'sqlDB' to uniformly manage read and write requests. The system designs independent function methods for each business data table, so as to ensure that the massive heterogeneous data generated in the scanning process can be safely and structurally processed and stored. At the same time of data writing, in order to meet the real-time monitoring requirements of security testers for task progress, the system constructs a terminal dynamic feedback mechanism, which extracts the latest status data from the database in real time and calls a special 'console' rendering function to convert the original data into a customized standardized terminal output format, which greatly improves the readability of complex logs and the tracking efficiency of vulnerability links under the command line interface. At the end of the life cycle of the scanning task, the system starts the final report delivery, and the core scheduler will fully evaluate the results from the database and serialize them into standard JSON data streams, Subsequently, the system introduces a predefined HTML template engine, reads the template file through a custom parsing function, uses data binding technology to accurately map JSON data and dynamically replace it into the corresponding placeholder of the HTML template, and finally automatically renders a visual security assessment report with clear structure, good interaction and high readability, so as to completely realize the project from bottom-level data persistence, middle-level real-time monitoring to top-level visual delivery. 
\section{Implementation}
The operation and empirical evaluation of this system rely on a highly standardized simulation security environment. The underlying operating system uses the Linux Kali 6.6.9-amd64 distribution, and the core operating environment is designated as Python 3.11.8. The target range used to verify the effectiveness of the system uses the BUUCTF online simulation platform and the vulhub open-source vulnerability Docker image collection to ensure the authenticity and vulnerability coverage of the test scenario. In the environment configuration and initialization phase, the system strictly requires that the Python version of the host computer should not be less than 3.10. Users only need to use the package manager to automatically and batch deploy all core third-party dependency libraries. In order to adapt to the interaction requirements in different security assessment scenarios, the system designed a flexible dual-mode startup mechanism: for the local test scenario that prefers the integration of efficient instruction operation and automated scripts, researchers can directly execute 'Python deepscan.py' in the root directory of the system to invoke the lightweight CLI. For scenarios requiring visual data presentation or remote server deployment, the system provides a web-based graphical startup scheme by running the 'Python start web.py' command in the root directory, the system will not only pull up the web service in the background and bind it to the whole network segment interface (0.0.0.0) by default to support cross-device remote access in the LAN, but also automatically invoke the local default browser and navigate to 'http://127.0.0.1:5000'. The core console provides a comprehensive test platform with both operational flexibility and architecture scalability for security evaluators.
\subsection{User Interface}
This tool offers two interaction methods: the terminal command line and a web-based graphical interface. When running in the system terminal, the pyfiglet library, which uses Python to generate ASCII art, is used to create a banner to enhance the user interface. Custom classes are employed to set the color and style of the text displayed on the interface. This code uses Python's pyfiglet library to create a banner with stylized text. It also employs custom classes and ANSI escape codes to control text color and formatting in the terminal. The web graphical interface was developed to align with modern user habits and enhance usability. Based on the Flask framework\cite{bednarz2025benchmarking}\cite{albesher2024observational}, a system-level web graphical interface was created, allowing users to access the tool directly via a browser. The web interface utilizes the Bootstrap framework to achieve a responsive layout, ensuring optimal display across different devices\cite{r2023web}.

\subsection{WebInfo Data}
WebInfo data includes IP addresses, geographic locations, webpage title fingerprints, server types, operating system information, and CDN and WAF identification. The following describes the specific implementation approach for each function.

(1)	Retrieving IP Addresses and Geolocation. When resolving IP addresses, the 'dns.resolver.Resolver' is used to perform DNS resolution, querying the A record of the domain name. An A record is a resource record type in the DNS protocol whose primary function is to map a domain name to an IPv4 address\cite{liu2023formal}. Resolution is then carried out using public DNS servers, such as 1.1.1.1 and 8.8.8.8. Invalid IP addresses, such as local loopback addresses, are filtered out. To obtain the geographic location, the 'geoip2' library is used to read the 'GeoLite2-City.mmdb' database file (which typically contains geographic location information associated with IP addresses). This allows us to query the city information for an IP address and retrieve the country, province, and city names\cite{livadariu2020accuracy}\cite{zhu2014survey}.

(2)	Operating System Information. In the automatic information collection and fingerprint identification module of the system, the underlying network discovery function integrates the 'Python nmap' library. By calling the locally deployed open-source nmap engine, it realizes the Python automatic encapsulation of network topology detection and security audit capabilities, and provides accurate port and service information for subsequent vulnerability verification. In the stage of feature extraction at the web level, the system effectively captures the target server type (server), page title, and multidimensional web fingerprint information through the 'info' method encapsulated in the 'webpage' class, and introduces the set operation mechanism to strictly deduplicate the multi-source detection data, so as to ensure the uniqueness and accuracy of fingerprint features. Aiming at the problem of "target IP distortion" caused by CDN widely existing in modern web architecture, the system designs a dual CDN recognition mechanism based on the combination of IP interval matching and ASN (autonomous system number) traceability\cite{zolfaghari2021content}\cite{wei2023cdn}\cite{ghaznavi2021content}: on the one hand, the system has built-in a large IP address segment feature library covering the mainstream CDN service providers at home and abroad, which can quickly determine whether the target IP belongs to the CDN edge node through the boundary matching algorithm, On the other hand, the system relies on the geoip2 database to query the ASN attribution of the target IP and cross compare it with the ASN list of known CDN providers\cite{butler2010survey}\cite{dimitropoulos2007relationships}, so as to accurately peel off the non real source IP and avoid invalid network overhead caused by subsequent vulnerability verification module on false targets. In addition, in the detection link of WAF, the system adopts a dynamic identification strategy combining active trigger and feature matching, that is, it injects malicious test payload into the target to induce the interception mechanism of WAF, and then captures and extracts the specific HTTP response features returned by WAF, and finally compares them with the locally constructed WAF fingerprint database in depth, so as to realize the accurate perception of the target defense environment and provide the basis for subsequent PoC scheduling.

\subsection{Port Scanning}
This program performs a port scan using fully established TCP connections and employs multithreading to increase the scanning speed\cite{everson2024survey}. It also includes a feature for identifying services based on port numbers. In service identification based on port, we first use the SIGNS list to match the response content with regular expressions to determine the service. If the match fails, we identify the service using the port: service dictionary defined in 'get-server'. In Python, the `concurrent.futures` module provides a high-level interface for asynchronously executing function calls. It includes two executors: `ThreadPoolExecutor` and `ProcessPoolExecutor`, which are used to manage and schedule the execution of threads and processes, respectively. Here, multithreading is used to implement concurrent requests.

\subsection{Directory Scanning}
Directory scanning involves four key aspects: dictionary generation, 404 false positive detection, response content verification, and concurrent requests. The following provides a detailed explanation of each of these four components:

(1)	Dictionary Generation. The Cartesian class is used to compute the Cartesian product, generating a large number of possible directory and filename combinations from a small set of dictionary files\cite{mcminn2004search}\cite{han2014password}\cite{nie2011survey}. The scan covers two categories: sensitive files and regular directory files. Sensitive files are identified by combining directory prefixes (e.g., /admin, /backup) with sensitive file extensions (e.g., .zip, .bak). Regular directory files are determined by using previously collected fingerprint information to identify file extensions, which are then combined with a standard directory dictionary to generate a more system-specific directory dictionary. 

(2)	404 False Positive Detection. The reason for performing 404 detection is that web administrators may customize 404 pages, which will return a 200 status code. Since we cannot rely on the status code to determine if a page is valid, the detection approach involves constructing invalid URLs to access 404 pages, recording their characteristics, and then comparing them with the 404 page characteristics during subsequent directory scans to improve detection accuracy. 

(3)	Response Content Verification. Use the -verify method to parse each response to determine whether the directory or file actually exists. This involves filtering out common error codes and unexpected content types, detecting common 404 page text, analyzing whether redirect targets are valid, and applying specific validation logic for different file types.

(4)	Concurrent Requests. Select an appropriate acceleration method based on the runtime operating system: use multithreading on Windows and asynchronous I/O on Linux/Unix to implement concurrent requests and accelerate the scanning speed. This approach was chosen because port scanning is inherently a highly concurrent I/O operation, specifically, a large number of concurrent TCP connection attempts. Asynchronous I/O can manage tens of thousands of concurrent connection attempts with relative ease and without incurring significant thread overhead. The uvloop library\cite{zhang2023faster} is a Python asynchronous I/O library based on Cython and libuv that enhances the performance of asyncio, however, it does not support the Windows platform and runs only on Unix/Linux systems\cite{sodian2022concurrency}. 

\subsection{FingerPrint/Server Identification}
This project employs a two-tier fingerprinting mechanism, utilizing the WebPage and Wappalyzer core classes to achieve service identification. At the basic identification layer, the WebPage class first uses the BeautifulSoup library to parse the HTML page of the target website\cite{olston2010web}, extracting the src attributes of all script tags, meta tag information, and HTTP response headers. It then performs rapid matching of the response headers and page content using regular expressions based on the feature rules defined in the `apps.txt` file, thereby achieving preliminary application identification. At the deep identification layer, the `Wappalyzer` class loads a more complex `apps.json` rule library capable of matching various service system fingerprints, such as CMS, forums, e-commerce systems, blogs, and more. Finally, the tool integrates the identification results from both layers to output complete web application fingerprint information.

\subsection{Proof Of Concept}
The key consideration for PoC verification is how to determine whether the target matches the PoC\cite{sejfia2024toward}\cite{zhou2019devign}\cite{fan2020c}\cite{li2018sysevr}\cite{uddin2025deep}\cite{zhang2023vuld}. Implementation approach: Since WebInfo data has already been collected, information such as open ports and application fingerprints is passed to the `check` function in the PoC script. This function is used to verify whether the conditions for executing the PoC are met, as well as to evaluate and handle the execution results\cite{li2018fuzzing}\cite{wang2024progress}\cite{manes2021art}. Additionally, multithreading is employed to run multiple POCs simultaneously for verification. The `check` function must be defined in accordance with the specific vulnerability mechanism, but its parameters are standardized. 

\subsection{Report Generation}
The system's report output primarily consists of three main components: real-time output on the terminal, generation of HTML-formatted reports for the corresponding targets, and writing report data to an SQLite database file. The following sections provide a detailed overview of the implementation details for each component\cite{liu2014survey}.

(1) Real-time terminal output. Each scan result is displayed using the format "Current Time | Function Module | Scan Result", and custom classes are used to set the text color and style to achieve a highlighted effect.

(2)	The section for generating HTML-formatted reports. Use custom HTML templates based on Vue.js and Bootstrap. Each generated HTML file is named "Deepscan-timestamp.html" and includes complete styling and interactive features.

(3)	Writing report data to a database file. This functionality is encapsulated in the sqldb.py file. The sqlite3 library is used to create an SQLite database, and the collected, scanned, and analyzed data is written to a DB file. The data can then be viewed using a database management tool such as Navicat. The web interface also provides a feature to view the DB file, enabling one-click access to scan results. 

\section{Evaluation}
\subsection{Effectiveness}
The vulnerability environments used for functional testing are the online CTF range BUUCTF and the Vulhub range. BUUCTF is a CTF competition and training platform that provides online reproductions of real-world challenges and vulnerability environments. The Vulhub range is an open-source collection of pre-built Docker environments containing vulnerabilities, designed for security researchers and educators. All vulnerabilities are real-world, and each comes with detailed documentation explaining the vulnerability and exploitation steps. Users can quickly access these virtual environments via Docker. Functional testing consists of the following modules, and all test cases have passed. Specific test cases are shown in Tables 5 and 6.

\vspace{-0.3em}
\begin{table}[H] % Set to uppercase H to force it to remain at its current position
  \centering
  \renewcommand{\arraystretch}{1.3}
  \renewcommand{\thetable}{5}
  \caption{Table of Basic Execution Test Cases}
  \label{tab:basic-test}
  
  % \begin{tabular}{C{1.8cm} C{1.8cm} C{3.8cm} C{4cm} C{1.8cm}}
  \begin{tabular}{@{} C{2.0cm} C{2.2cm} C{4.2cm} C{4.5cm} C{1.4cm} @{}}
    \toprule
    \textbf{Test Case} & \textbf{Prerequisites} & \textbf{Test Steps} & \textbf{Expected Result} & \textbf{Passed} \\
    \midrule
    Parameter Processing & None & Process user-input parameters, normalize parameters & Processed parameters ensure normal function execution & $\checkmark$ \\
    Start Execution & None & User launches the main program, scans specified targets & Scans normally, stable execution & $\checkmark$ \\
    Function Call & None & After the main program runs, automatically calls various function modules & Standardized output of execution results for each function & $\checkmark$ \\
    \bottomrule
  \end{tabular}
\end{table}

\begin{table}[H]
  \centering
  \renewcommand{\arraystretch}{1.3}
  \renewcommand{\thetable}{6}
  \caption{Functional Execution Test Cases Table}
  \label{tab:func-exec-test}
  
  \begin{tabular}{@{} C{2.0cm} C{2.2cm} C{4.2cm} C{4.5cm} C{1.4cm} @{}}
    \toprule
    \textbf{Test Case} & \textbf{Prerequisites} & \textbf{Test Steps} & \textbf{Expected Result} & \textbf{Passed} \\
    \midrule
    Liveness Check & None & \makecell[c]{Can correctly determine\\whether the target is alive} & \makecell[c]{Can properly identify\\the target's status} & $\checkmark$ \\
    
    DNS Resolution & None & \makecell[c]{1. Determine whether user input\\is domain name or IP\\2. If domain name format,\\perform DNS resolution} & \makecell[c]{Correctly returns IP address\\regardless of input format} & $\checkmark$ \\
    
    IP Geolocation & \makecell[c]{Pre-configured\\mmdb database} & \makecell[c]{Uses geoip library to query\\IP location database} & \makecell[c]{Returns geolocation\\for public IPs} & $\checkmark$ \\
    
    WebInfo & None & Records website characteristics & \makecell[c]{Correctly identifies\\website characteristics} & $\checkmark$ \\
    
    WAF Detection & \makecell[c]{Uses signature\\database} & \makecell[c]{Send malicious payload to\\trigger WAF, record signature\\matches in database} & \makecell[c]{Triggers WAF and\\returns results} & $\checkmark$ \\
    
    CDN Detection & \makecell[c]{Uses ASN\\database} & \makecell[c]{Determines if IP belongs to\\common IP ranges and ASNs} & \makecell[c]{Outputs CDN\\determination result} & $\checkmark$ \\
    
    OS Detection & Uses nmap & Invokes nmap to identify OS & Outputs identification result & $\checkmark$ \\
    
    Side-Site Detection & None & Uses API interface for detection & Returns results normally & $\checkmark$ \\
    
    PoC Detection & None & \makecell[c]{Based on collected fingerprints,\\pass PoC script to determine\\if execution conditions are met} & \makecell[c]{If PoC passes, display\\specific vulnerabilities\\in target} & $\checkmark$ \\
    \bottomrule
  \end{tabular}
\end{table}

\subsection{Robustness}
This tool utilizes a wide range of Python third-party library functions, however, some of these libraries do not support operation on Windows. For example, uvloop is a fast, scalable event loop for Python designed to replace the default event loop implementation in Python's standard library asyncio and significantly improve the performance of asynchronous I/O operations. Therefore, for the directory scanning functionality, a multi-threaded Python thread pool is used when running on Windows, while an asynchronous I/O strategy is employed on Linux. This ensures that the tool runs normally on both Windows and Linux terminals. Empirical results confirm that the proposed fingerprint-driven mechanism successfully bridges the semantic gap between reconnaissance and utilization, and achieves efficiency and accuracy with extremely low network overhead.
\section{Conclusion}
This paper combines network scanning, reconnaissance, and penetration testing techniques to design and develop a lightweight yet powerful cybersecurity reconnaissance tool called Deepscan. The tool runs smoothly on various operating systems and has been optimized for performance across different operating system environments. It implements features such as directory scanning, port scanning, and fingerprinting. Deepscan involves a wide range of technical stacks, including network programming, multithreading, asynchronous programming, page parsing, data analysis, and database operations. Compared to existing open-source scanning systems, Deepscan's advantages lie in its scanning depth, adaptability, and efficiency. It can automatically select the optimal technical solution based on the runtime environment, delivering reliable performance even with limited resources. It adopts a modular design with low coupling between functional components, facilitating easier expansion. DeepScan implements the entire workflow from reconnaissance to vulnerability verification, lowering the technical barrier for security assessments and making it easy for non-specialists to learn and use. DeepScan has been validated in various areas, including website security assessments and vulnerability discovery, confirming its practicality and effectiveness.

\section*{Acknowledgments}
We thank Meili Zhao for her valuable advice. AI-based tools are used for language polishing during manuscript preparation.

%Bibliography
\bibliography{references}  

\end{document}